# TweetIT- Analyzing Topics for Twitter Users to garner Maximum Attention

Dhanasekar Sundararaman
SSN College of Engineering
Chennai, Tamilnadu
India
dhanasekar312213@gmail.com

Priya Arora
Texas A&M University
Texas
United States
arora.priya4172@gmail.com

Vishwanath Seshagiri
College of Engineering, Guindy
Chennai, Tamilnadu
India
vishwa.nath@outlook.com

## ABSTRACT
Twitter, a microblogging service, is today's most popular platform for communication in the form of short text messages, called Tweets. Users use Twitter to publish their content either for expressing concerns on information news or views on daily conversations. When this expression emerges, they are experienced by the worldwide distribution network of users and not only by the interlocutor(s). Depending upon the impact of the tweet in the form of the likes, retweets and percentage of followers increases for the user considering a window of time frame, we compute attention factor for each tweet for the selected user profiles. This factor is used to select the top 1000 Tweets, from each user profile, to form a document. Topic modelling is then applied to this document to determine the intent of the user behind the Tweets. After topics are modeled, the similarity is determined between the BBC news data-set containing the modeled topic, and the user document under evaluation. Finally, we determine the top words for a user which would enable us to find the topics which garnered attention and has been posted recently. The experiment is performed using more than 1.1M Tweets from around 500 Twitter profiles spanning Politics, Entertainment, Sports etc. and hundreds of BBC news articles. The results show that our analysis is efficient enough to enable us to find the topics which would act as a suggestion for users to get higher popularity rating for the user in the future.

## CCS Concepts
• **Information systems**→**Database management system engines** • **Computing methodologies**→**Massively parallel and high-performance simulations.** This is just an example; please use the correct category and subject descriptors for your submission. The ACM Computing Classification Scheme:

## 1. INTRODUCTION
Social networks like Twitter, Facebook etc. are becoming an essential part of everyday life. Being a microblogging platform, Twitter has gained so much of attention in the past few years since it allows users to post short text messages, called Tweets on their timeline. These Tweets range from news articles to gossips but within a limit of 140 characters. It enables interested parties to follow individual user's views on an event or a topic in real-time[1]. Users tend to follow people with whom they share common interests or who they want to know more about like their favourite celebrities etc. Hence, people tend to post Tweets that can bring maximum attention to their engaged followers as well as which can fetch more followers to the profile. An analysis on this is imperative as it tends to reveal the topics a particular user has posted, which fetched a lot of attention enabling them to do the same in the future.

Now after developing an efficient service for determining the potential suggested topics, a user can enhance his follower list. This list is expanded gradually depending upon the topic chosen for the tweet which, in turn, leads to getting most popularity ratings. The newly added followers would have shared the same interests like the topic of the tweet, or the topic trending in the market. For example, if the user is a healthcare consultant and tweets are highly focussed on health-insurance, health-benefits, meditation and medical-equipments etc., the followers of that user would have interest in this domain. However, if the user is a popular celebrity or a politician, the topic of the tweet may span various topics ranging from entertainment to politics. Hence, with the liking in several domains, the user gets increment in his follower list within a time frame. Though this seems obvious, what if there is a way to find the topics, among several interested once, which a user can focus the most so that his follower list could be increased; and this could lead to getting most popularity ratings for that user profile based upon the tweet. To get this rating, we propose a topic-modeling based twitter service that would help in determining the top words that a user could use for enhancing the follower's list and hence popularity ratings.

To summarise,

- We take the last 3000 tweets from Twitter API of several users (over 500) of various domains like entertainment, politics, sports. Further, we analyse, pre-process and mine the data.

- For each user, we select top 1000 tweets based on the attention factor of the tweets. This attention factor is determined by the number of likes on the tweet and retweet count, on a topic, within a time-frame window.

- Then we apply the topic modelling on each user document containing top 1000 tweets. BBC news data-set is then analysed to find the topic and compared with the document of each user. The similarity between the user's tweet top-document and the BBC data-set suggests us the topic that a user should tweet upon to get the maximum popularity ratings.

The paper is organised as follows. Section 2 deals with the literature survey about the Twitter as a major microblogging platform in social media. It also highlights scenarios where user modelling can be done on the Twitter API. Section 3 talks about the collection of Twitter and BBC news dataset and the filters employed to extract Tweets from the user efficiently. Section 4 describes the profile modeling in which the most representative terms are found and the similarity measures employed and the accuracy with which proposed study can be used for the suggestion of the topics to the user. Section 5 explains the process behind our research work while section 6 highlights the results of the process. We conclude finally with Section 7 reflecting on the shortcomings with the future vision of this work.

## 2. RELATED WORKS

The social microblogging services/platform such as Twitter helps in engaging people by letting them publish about their thoughts ranging from entertainment to politics. Millions of people, now a day is part of this microblogging platform. Everyone uses Twitter for conveying their views and to follow like-minded people who post such views. Due to such a large user-base, this platform has gained lots of attention from the research point of view. Hence, to study the work already done in this direction, we provide the literature survey that is most closely related to our work. The work done by [2,3] describes the notion of Twitter being a microblogging website and how it impacts the society. It also hypothesises the differences between a users' network connection structure by suggesting three types of unique user activities - information sharing, information seeking and social activity. Likewise, work done by [4,5,6] provides us with a background as to how the tweets can be handled for such a large user-base platform services. Moreover, it gives us the strategies as to how tweets can be analysed. Similarly, the work done by Liu et al. [7] gives us several tools for mining the data obtained from various data-sets.

Apart from the study of Twitter, a micro-blogging service and how the tweets can be analysed, much of the work has been done in studying the pattern of tweets, like the work was done by Krishnamurthy et al. [8] perform a descriptive analysis of the Twitter network. It suggests that frequent updates might be correlated with high overlap between friends and followers. Similarly, Cha et al. [9] study the impact of mentions, retweets and on the dynamics of on topics and time. Apart from the impact of the users' tweet on the community, there has been work done as to how the expressed tweet content influence the society as studied by Cataldi et al. [10]. The same kind of work has been done by Rui et al. as described in work [11]. In another work, Guo et al. [12] the patterns of user tweets in several online communities, such as a blog system, a question-answering social network and a social bookmark-sharing network. They find that the rank order distribution of user postings follows stretched exponential distribution. There is one study which indirectly relates to our hypothesis, as mentioned above, by Huberman et al. [13] which says that analysis of data from YouTube exhibited in crowdsourcing exhibits a strong positive dependence on attention, measured by the number of downloads. They find that the lack of interest leads to a decrease in the number of uploaded videos and a hence a drop in productivity.

There is one more parameter to determine the popularity of a Twitter account, and that is the follower counts. Many studies have been done as what makes a Twitter follower count and the factors that impact its increment as done by [14,15, 16, 17]. Depending upon the discussion on the topics based upon the recency and variety of topics happen on Twitter, user profiles generated from Twitter posts promise to be beneficial for other applications on the social web as well. However, automatically inferring the semantic meaning of Twitter posts is a non-trivial problem as mentioned in work [18].

## 3. DATA COLLECTION AND PREPROCESSING

In this section, first of all, we describe the data-set obtained from Twitter API and then the process using which tweets are collected related to several domains - sports, politics, education, entertainment, and philosophy. Then, we explain about the BBC dataset and finally about preprocessing these data.

### 3.1 Twitter Data Collection

The process used for filtering out the data-set is as follows:- First of all, a large number of Twitter profiles are selected, each contains a large number of tweets. A popularity filter is employed to find the Twitter profiles having a minimum threshold of followers (several thousand). Second, a Language filter is applied to extract the words used for tweeting by the users. Language filter provides consistency across geographically distributed domains for the same user profile where the same user profile is perceived differently due to the usage of different language depending on the region. Further, a domain filter is applied to extract user profiles by comparing it with the words or topics extracted from the BBC news dataset. To perform this experiment, we have extracted approximately 3000 user profiles of renowned people among various domains. At an average of 120 characters per tweet, our dataset comprises 1.1M tweets with 130M characters approximately.

### 3.2 News Data Collection

The news articles data is collected from the news provider, BBC available publicly online. It spans from various topics like politics, sports, entertainment, regional news, religion and current trends. These news articles were carefully chosen to include the interests of that of various Twitter profiles to avoid any bias. Hundreds of such news articles are preprocessed to remove any abbreviations and acronyms. The preprocessed text is then aggregated into a single text document as a collection of news articles. This forms the query document, that is the document that is queried on different Twitter profiles for similarity.

### 3.3 Preprocessing

The first step in pre-processing is to transform the data to get an efficient format so that the top influential words of the tweets for each user profile can be extracted. Here, preprocessing essentially deals with filtering out non-required words from a user's profile. The non-required words are selected by sorting the bag of words selected for each user profile. This sorting is done by the number of likes and increment in follower count to that user profile. To filter out the acronyms in a tweet, we pre-process the data by using acronym dictionary by converting acronyms to their full forms. For example, "omg" is translated to "Oh My God". During pre-processing, words containing symbols such as '@,' '#' and numerics are removed. The 1.1M such preprocessed tweets are carefully documented based on their domains.

To process the data using our proposed methodology, it is required for the Tweets not be just arranged as stand-alone strings but rather as user profile text documents. Hence 1000 such selected tweets from each user profile are merged to form one single profile document, which collectively makes up an exhaustive dataset.

## 4. PROFILE MODELING AND SIMILARITY

In this section, first, we discuss the profile modeling by determining the most representative words of that particular user profile (Section 4.1), and then we determine the similarity the user's document (Twitter profile) and the BBC data-set using suitable distance metric (Section 4.2).

### 4.1 User Profile Modeling

User profile modelling refers to the top influential words of a user profile, by top influential we mean the most important words that gain the maximum number of likes with the increment in the number of followers to that user. As already defined, this importance is a measure by the number of likes and increment in the number of followers to the user profile based on the tweet. For determining the same, we compute TF-IDF between the documents generated using user's profile with the one obtained using BBC data-set for the same influential word. TF-IDF is the feature selection approach to determine the highly representative words of a document by ignoring the words that don't convey any meaning to the content like stop words.

**Table 1. TF-IDF of few sample terms in @realDonaldTrump**

| realDonaldTrump | TF-IDF |
|---|---|
| taxes | >0 |
| election | >0 |
| washington | >0 |
| reforms | >0 |
| immigrant | >0 |

**Table 2. TF-IDF of few sample terms in @katyperry**

| katyperry | TF-IDF |
|---|---|
| taxes | 0 |
| election | 0 |
| washington | >0 |
| reforms | 0 |
| immigrant | 0 |

Table 1 shows the TF-IDF of sample terms in @realDonaldTrump profile. These words were completely picked at random. The TF-IDF scores of these words are greater than 0, which implies that these words have been used a considerable number of times and are important or characteristic of that Twitter profile. We then see the TF-IDF scores of the same words in another contrasting Twitter profile. Table 2 shows that only some of the words enlisted (in fact only one out of five) have a score, the rest are zeros. This indicates that the first Twitter profile, being a one that of politician involves a lot of terms related to government and country. The second profile, on the other hand, belongs to the entertainment industry, and obviously, the words that relate to the notion of government and politics has a score of zero. When the similarity between these two profiles is analysed, they were found to be far less than that of two profiles involving political domain.

## 4.2 Profile Similarity

Now that the top influential words in a user profile are identified using TF-IDF scores, the similarity between the user profile (Twitter profile) document with the BBC data-set document can be compared. Various distance metrics exist to compute this similarity such as Cosine, Euclidean, and Manhattan, etc. However, Cosine distance is the appropriate choice [19] for this work since it takes care of the bias caused due to the difference in the document lengths. This similarity is computed using TF-IDF which computes inner product of two vectors generated for the two documents (sum of the pairwise multiplied elements) which are further divided by the product of their vector lengths. The resultant score lies in the range of 0 and 1. Hence, the similarity score is obtained by subtracting this value from 1.

$$\text{Cosine distance} = (d1d2)/\|d1\|\|d2\|$$

Where d1 and d2 are the vectors of the same length.

$$\text{Cosine similarity} = 1 - (d1d2)/\|d1\|\|d2\|$$

While calculating the similarity between the user profile document and BBC dataset generated querydocument, the top terms contained in the query document alone are taken. The TF-IDF scores corresponding to the top words present in the user tweet is used for cosine distance computation. That is, the influential words present in the user profile are searched in the document generated using BBC dataset for the same word. Next, the TF-IDF scores of these two documents are compared for finding the similarity.

## 5. TweetIT – The Process

The process behind TweetIT can be broken down into three steps.

## 5.1 Gathering Popular Tweets

For modelling a person's popularity, one has to consider the reach which the person has amongst the Twitterati. Twitter provides us with a quantitative measure in the form of Likes and Retweets for every tweet, which can be used to gauge the person's ken about a certain topic or event. We combine these measures to calculate the attention factor, which is an average of some retweets and likes.

Attention_Factor = 0.5 * No of Likes + 0.5 * No of Retweets

This Attention_Factor is calculated for all the Tweets, and the top 1000 Tweets for each profile is obtained and stored in a separate document. Once these tweets are gathered, the time that has elapsed since the last posting this tweet is obtained and sorted from latest to oldest Tweet. This is done so that the tweet which is most recent and has highest attention_factor appear at the top. Using these newly acquired Tweets, Topic Modelling (Section 4.1) is performed to identify the topics on which a user is Tweeting.

## 5.2 Analyzing Similarity

After gathering recent popular Tweets for each of the 500 Twitter profiles, we then analyse the similarity of these Tweets with the BBC news articles. A percentage score of similarity is obtained for each profile with every news articles. The similarity is achieved by the presence of words/ topics present in the Tweets that are also present in the news articles

For each profile, we compare its similarity using the formulas described in section 4.2 with all the latest news articles and find the most similar user and the second most similar user and so on.

## 5.3 Analyzing Topics

After finding similarity between the news articles and the various Twitter profiles, we then analyse the words or in more technical form topics, that caused the similarity in the first place. We use Influential Term Metric, proposed in [20] that finds the topics that have high influence individually in the Twitter profile, news data and together as a whole.

The topics may be biased based on the nature of the dataset. If the news articles contained in the majority of sports, the most similar profiles could be that of a sportsperson, and the top words could be that related to sports. A glimpse of the results obtained through BBC news articles and Twitter profiles is shown in Tables 3 and 4 in the next section.

## 6. RESULTS

**Table 3. Similarity Ranking of Twitter profiles to BBC News**

| QUERY DOCUMENT | TWITTER PROFILE | RANK |
|---|---|---|
| BBC NEWS DATASET | faisalislam | 1 |
| | jamesrbuk | 2 |
| | benfenton | 3 |
| | robfordmancs | 4 |
| | paullewismoney | 5 |
| | DouglasCarswell | 6 |
| | montie | 7 |
| | PatrickStrud | 8 |
| | JolyonMaugham | 9 |
| | IanDunt | 10 |

For finding the result, we combined 300 BBC news articles from various domains and used it as our query document. It was a collection of all the latest trending news from across the globe. Similarly, top 1000 Tweets of each user were sorted by the posted date, and stored in the separate document, which was used to compare against the query document. Based on similarity explain in 4.2, we have ranked the profiles Table 3 shows the top 10 profiles, whose tweets are similar to topics trending in BBC dataset. When we get all news articles from other genres, it can show how the ranking varies based on the content of news. This goes on to show that, even if personal tweets could be ranked, there needs to be a correlation with the trending topics, to propagate your opinions better.

Apart from finding similarity between a news article and Twitter profiles, we go a notch higher and find the topics that are influential [20]. Five of the top such topics, between each Twitter profile, and BBC News is shown in Table 4. Top 5 such topics of @faisalislam are 'labour, government, chancellor, market, tories' which have been recently posted and garner maximum attention from the Twitteratti.

Thus our system enables one to find many such topics for each Twitter profile by latest news sources, and propagate their views across depths of social media.

**Table 4. Top words in most similar Twitter profiles**

| NAME | TOPWORDS |
|---|---|
| faisalislam | labour |
| | government |
| | chancellor |
| | market |
| | tories |
| jamesrbuk | people |
| | britain |
| | premiership |
| | public |
| | secretary |
| benfenton | mobile |
| | election |
| | british |
| | england |
| | digital |
| robfordmancs | voters |
| | research |
| | growth |
| | ireland |
| | economic |
| paullewismoney | pensions |
| | customers |
| | compensation |
| | ryanair |
| | company |

## 7. CONCLUSION

Thus, we were able to determine topics of Twitter profiles that were posted recently, has attained widespread recognition and found in that of news headlines. This enables the user to retrospect on topics that garnered more attention than average, which in turn helps to grow the followers. The system can work to find such topics for any Twitter profile and a query document. The query document can be a news article, advertisement or any text for which related Twitter profiles is to be found. There are certain drawbacks in this process such as the failure to detect semantic words – words with the same meaning but in different forms. In the future, we would use dictionary such as Wordnet to eliminate redundancies caused by such misinterpretations.

## 8. REFERENCES

[1] Schonfeld, E.: Mining the thought stream. TechCrunch Weblog Article (2009), http://techcrunch.com/2009/02/15/mining-the-thought-stream/

[2] Java Akshay, Xiaodan Song, Tim Finin, Belle Tseng. Why We Twitter: Understanding Microblogging Usage and Communities. Springer, 56-65, 2007..

[3] Mor Naaman, Jeffrey Boase, Chih Hui Lai. Is it really about me?: Message content in social awareness streams. 189-192, 2010.

[4] Aditi Gupta, Ponnurangam Kumaraguru. Credibility ranking of tweets during high impact events. 978-1-4503-1236-3, 2012.

[5] Danah Boyd, Scott Golder, Gilad Lotan. Tweet, Tweet, Retweet: Conversational Aspects of Retweeting on Twitter. 1530-1605, 2010

[6] Honeycutt, C., & Herring, S. Beyond microblogging: Conversation and collaboration via Twitter. In Proc. HICSS '09. IEEE Press (2009).

[7] Liu, B.: Web data mining; Exploring hyperlinks, contents, and usage data. Springer, Heidelberg (2006) - how to mine data.

[8] Krishnamurthy, B., Gill, P., and Arlitt, M. A few chirps about twitter. In Proc. WOSP '08. ACM Press (2008)

[9] Meeyoung Cha, Hamed Haddadi, Fabrício Benevenuto, Krishna P. Gummadi. In Proc. ICWSM (2010).

[10] Mario Cataldi, Luigi Di Caro, Claudio Schifanella; Emerging topic detection on Twitter based on temporal and social terms evaluation. In Proc MDMKDD'10 2010.

[11] Huaxia Rui, Andrew Whinston; Information or attention? An empirical study of user contribution on Twitter. In Proc. Information Systems and e-Business Management 2012.

[12] Guo, Lei, et al. "Analyzing patterns of user content generation in online social networks." Proceedings of the 15th ACM SIGKDD international conference on Knowledge discovery and data mining. ACM, 2009.

[13] Bernardo A. Huberman, Daniel M. Romero, Fang Wu; Crowdsourcing, attention and productivity. In Proc KDD, 2009.

[14] Gianluca Stringhini, Gang Wang, Manuel Egele, Christopher Kruegel, Giovanni Vigna, Haitao Zheng, Ben Y. Zhao; Follow the green: growth and dynamics in twitter follower markets. In Proc IMC 2013.

[15] Haewoon Kwak, Changhyun Lee, Hosung Park, Sue Moon; What is Twitter, a social network or a news media?. In Proc WWW 2010.

[16] Meeyoung Cha, Hamed Haddadi, Fabrício Benevenuto, Krishna P. Gummadi; Measuring user influence in Twitter: The million follower fallacy. In Proc ICWSM 2010.

[17] Chunliang Lu, Wai Lam, Yingxiao Zhang; Twitter User Modeling and Tweets Recommendation Based on Wikipedia Concept Graph. In Proc AAAI 2009.

[18] Fabian Abel, Qi Gao, Geert-Jan Houben, Ke Tao; Semantic Enrichment of Twitter Posts for User Profile Construction on the Social Web. In Proc ESWC 2011.

[19] Huang, Anna. "Similarity measures for text document clustering." Proceedings of the sixth new zealand computer science research student conference (NZCSRSC2008), Christchurch, New Zealand. 2008.

[20] Sundararaman, Dhanasekar, and Sudharshan Srinivasan. "Twigraph: Discovering and Visualizing Influential Words Between Twitter Profiles." International Conference on Social Informatics. Springer, Cham, 2017.

# Columns on Last Page Should Be Made As Close As Possible to Equal Length

## Authors' background

| Your Name | Title* | Research Field | Personal website |
|---|---|---|---|
|  |  |  |  |
|  |  |  |  |

*This form helps us to understand your paper better, **the form itself will not be published.**

*Title can be chosen from: master student, Phd candidate, assistant professor, lecture, senior lecture, associate professor, full professor